# Measurement of the atmospheric primary aberrations by 4-aperture DIMM

**Ramin Shomali,**[1] **Sadollah Nasiri,**[1,2] **Ahmad Darudi,**[1,3]

[1]Physics Department, Zanjan University, Zanjan 45195-313, Iran.
[2]Institute for Advanced Studies in Basic Science (IASBS), Zanjan 45195-1159, Iran
[3]Lund Observatory, Lund, Sweden

E-mail: shomali@znu.ac.ir

**Abstract.** The present paper investigates and discusses the ability of the Hartmann test with 4-aperture DIMM to measure the atmospheric primary aberrations which, in turn, can be used for calculation of the atmospheric coherence time. Through performing numerical simulations, we show that the 4-aperture DIMM is able to measure the defocus and astigmatism terms correctly while its results are not reliable for the coma. The most important limitation in the measurement of the primary aberrations by 4-aperture DIMM is the centroid displacements of the spots which are caused by the higher order aberrations. This effect is negligible in calculating of the defocus and astigmatisms, while, it cannot be ignored in the calculation of the coma.

## 1. Introduction

Evidence shows that atmospheric turbulence functions as the most important limitation in the astronomical high resolution imaging [1-3]. For the purpose of quantitative measurement of such turbulence above the telescope, several methods have been proposed by the astronomers up to now [4]. Of these, Differential Image Motion Monitor (DIMM) has proved to be the most common one [5-8].

Differential Image Motion Monitor consists of the optical systems in which the light that is passing through the two widely separated small apertures is separated by means of a small wedge prior to its falling on a CCD [6, 7].

The light from a single star illuminates each sub-aperture with a different column of air in front of which the turbulence induces phase fluctuations. These phase fluctuations, in turn, produce random motion for each sub-image. While the telescope vibrations affect each image in the same manner, the existing turbulence induces random differential motions in the sub-images. Thus, variations in the image separations can be used for obtaining a quantitative estimate of the turbulence [8]. For the Kolmogrov turbulence at the near field approximation, the longitudinal and transverse variances of the differential image motion for the two sub-apertures are related to the Fried parameter [6, 8]. As to this, the measurement of the longitudinal and transverse variances of the differential image motion can be used to estimate the Fried parameter in the case of two sub-apertures.

In optical testing, a very common method for testing the quality of the optical components is the Hartmann method. To test the shape of the optical surfaces, this method is frequently employed through using a screen with many apertures. The simplest Hartmann test can be performed by means of the Hartmann screen with four apertures for measuring some primary aberrations [9-11]. In the Hartmann test with four apertures screen, the apertures are located on the corners of a square. In this testing method, measurement of some primary aberrations is both possible and applicable for the alignment of the optical system, measurement of the focus errors, detections of decenterings, measurement of the astigmatism and for the coma too [9].

In this paper, we present a modification to the DIMM method. With the inclusion of two additional apertures to the DIMM, this modified method not only makes it possible to estimate the Fried parameter but, more importantly, gives a way to the determination of the three primary aberrations of the atmosphere: defocusing, astigmatism with axis at 0 or 90, and astigmatism with axis at $\pm 45$.

Parts of the evidence for such claims come from a study done by Tokovinin and his coworkers in 2008 [12]. They showed that the atmospheric coherence time could be calculated by measuring and processing atmospheric defocus fluctuations. For measuring the atmospheric defocus, they transformed stellar point images into the ring image by increasing the central obstruction and adding spherical aberration to the defocus aberration through the conic lenses. It seems that use of a 4-aperture DIMM for the atmospheric defocus measurement is much easier than the above method.

Focusing on the study aims, the section that immediately follows provides a theoretical account of the Hartmann test including a screen with four apertures and its application in calculating the primary aberrations. In section 3, the article proceeds toward describing the simulation of the 4-aperture DIMM and discuss the ability of this instrument for the measurement of atmospheric primary aberration. Finally, important concluding remarks are given in the last section.

**2. Hartmann test with 4-aperture screen**

Here a short review on the theoretical concept of the primary aberrations measurement by the four apertures Hartmann test is given.

Let us assume at the Hartmann screen, each aperture is located in one corner of a square with side $d'$ (figure1). The aberrations of the distorted wave-front at 4-aperture screen are defined by

$$W(x,y) = Bx + Cy + D(x^2 + y^2) + E(x^2 - y^2) + Fxy + G(x^2 + y^2 - d'^2)y + H(x^2 + y^2 - d'^2)x,$$

(1)

where B and C are the tilts about the y and x axis, D is the defocusing, E and F are the astigmatisms with the axis at $0$ or $90$ and at $\pm 45$, G and H are the comas along the y and x axis, respectively.

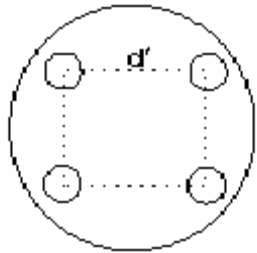

Figure 1. Four apertures Hartmann screen configuration.

As Salas-Peimbert et. al [11] pointed out, by this definition for the wave-front aberrations the centroids of these four spots (the average coordinates of the four spots) are not shifted by defocusing, astigmatism and coma terms. However, they are shifted only by the two tilts and the configuration of the system of four spots depends on the D, E, F, G, H coefficients while the global position is dependent upon on the coefficients of B and C.

The x and y components of the transverse aberrations are given by

$$\frac{\partial W(x,y)}{\partial x} = \frac{ta_x}{F_{tel}} = B + 2Dx + 2Ex + Fy + 2Gxy + H(3x^2 + y^2 - d'^2), \qquad (2)$$

and

$$\frac{\partial W(x,y)}{\partial y} = \frac{ta_y}{F_{tel}} = C + 2Dy - 2Ey + Fx + G(x^2 + 3y^2 - d'^2) + 2Hxy, \qquad (3)$$

where $F_{tel}$ is the 4-aperture screen distance from CCD, $\frac{ta_x}{F_{tel}}$ and $\frac{ta_y}{F_{tel}}$ are the angular transverse aberrations measured from their corresponding ideal positions.

Because in the Hartmann test with the four apertures, the presence of the two tilts in x and y directions displaces the centeroid of these spots from the ideal spots position, thus, we can calculate tilt coefficients by deviation of the centroid from its ideal position(see figure 2).

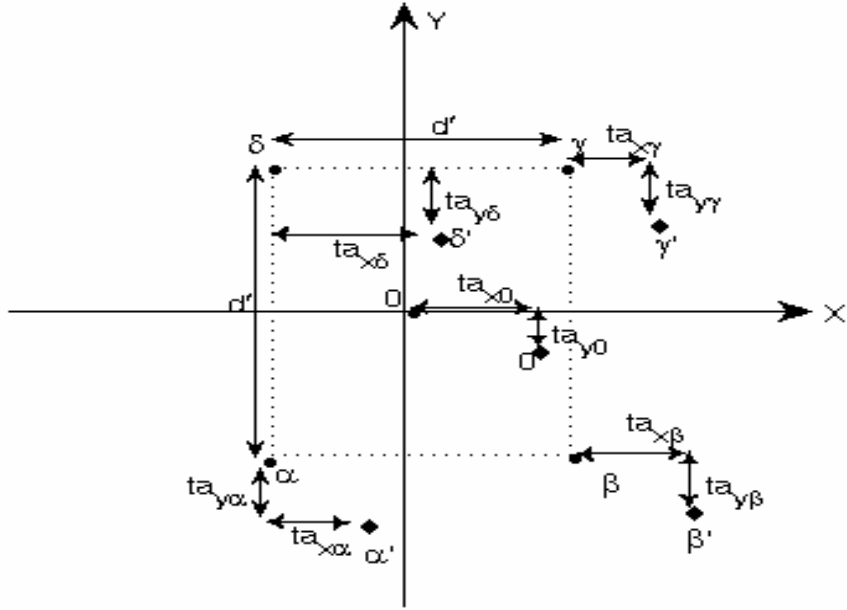

Figure 2. Ideal spots (circles) and real spots (diamond).

$$B = \frac{ta_{y\alpha} + ta_{y\beta} + ta_{y\gamma} + ta_{y\delta}}{4F_{tel}}, \qquad (4)$$

$$C = \frac{ta_{x\alpha} + ta_{x\beta} + ta_{x\gamma} + ta_{x\delta}}{4F_{tel}}, \qquad (5)$$

where $\alpha$, $\beta$, $\gamma$ and $\delta$ correspond to each of the apertures. The defocus, astigmatism and Coma coefficients (D, E, F, G, H) can be thus calculated by using Eqs.2-5.

Having obtained B and C from Eqs. 4 and 5, one can use them in Eqs. 2 and 3 to determine the aberration coefficients. The results come below

$$D = -\frac{(ta_{x\alpha} - ta_{x\beta}) - (ta_{x\gamma} - ta_{x\delta}) + (ta_{y\alpha} + ta_{y\beta}) - (ta_{y\gamma} + ta_{y\delta})}{8F_{tel}d'}. \qquad (6)$$

$$E = -\frac{(ta_{x\alpha} - ta_{x\beta}) - (ta_{x\gamma} - ta_{x\delta}) - (ta_{y\alpha} + ta_{y\beta}) + (ta_{y\gamma} + ta_{y\delta})}{8F_{tel}d'}. \qquad (7)$$

$$F = -\frac{(ta_{x\alpha} + ta_{x\beta}) - (ta_{x\gamma} + ta_{x\delta}) + (ta_{y\alpha} - ta_{y\beta}) - (ta_{y\gamma} - ta_{y\delta})}{4F_{tel}d'}. \qquad (8)$$

$$G = -\frac{(ta_{x\alpha} - ta_{x\beta}) + (ta_{x\gamma} - ta_{x\delta})}{2F_{tel}d'^2}. \tag{9}$$

$$H = -\frac{(ta_{y\alpha} - ta_{y\beta}) + (ta_{y\gamma} - ta_{y\delta})}{2F_{tel}d'^2}. \tag{10}$$

**3. Simulation**

To test the ability of the 4-aperture DIMM in measuring the atmospheric primary aberrations, we performed a numerical simulation the explanation of which comes below.

*3.1 Simulation of the four spots images at the telescope focal plane*

First, let us suppose that a single star light has a perturbed phase, $\varphi$, and a uniform illumination, $A$, in front of the telescope aperture. We already know that the complex wave function on telescope aperture is

$$U_i = A\exp(-i\varphi). \tag{11}$$

And, for a telescope with pupil function $P$, one can see

$$P(x, y) = \begin{cases} 1 & in\_side\_the\_aperture \\ 0 & otherwise \end{cases}. \tag{12}$$

Then, we know that the complex wave function and the intensity distribution at the focal plane of the telescope are [13]

$$U_f = FFT(PU_i) = FFT(PA\exp(-i\varphi)), \tag{13}$$

and

$$I_f = |U_f|^2, \tag{14}$$

where FFT stands for Fast Fourier transform. The simulation arrangement consists of the monochromatic light ($\lambda = 0.5\mu m$) that impinges on the telescope aperture with focal length $F_{tel} = 2.8m$, and aperture diameter of $D_{tel} = 28cm$.

In the aperture plane, we simulated a circular pupil with diameter 280 pixels, located at the center of a sampled rectangular matrix with $400 \times 400$ pixel resolution and $1mm \times 1mm$ pixel size. The observation plane was placed at the focal plane of a telescope with $400 \times 400$ pixel resolution and $3.5\mu m \times 3.5\mu m$ pixel size.

For simulation of spot images in 4-aperture DIMM, as illustrated in figure 3, we use the following pupil functions for each aperture with radius $R = 3cm$.

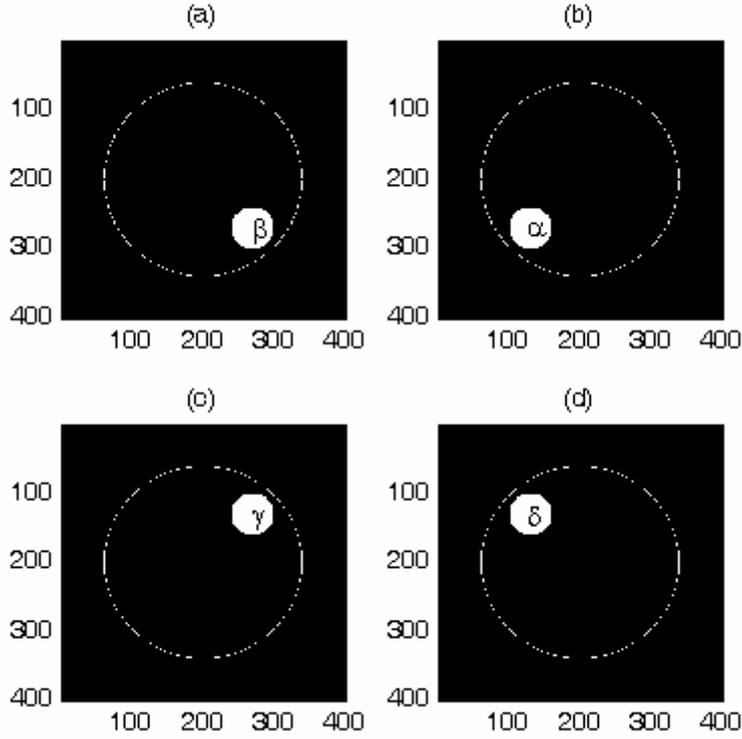

Figure 3. Configuration for the pupil function (a) the $\beta$ aperture, (b) the $\alpha$ aperture, (c) the $\gamma$ aperture and (d) the $\delta$ aperture. The dotted line shows the telescope aperture boundary.

$$P_\alpha(x, y) = \begin{cases} 1 & (x+70)^2 + (y+70)^2 \leq R^2 \\ 0 & otherwise \end{cases}, \qquad (15)$$

$$P_\beta(x, y) = \begin{cases} 1 & (x-70)^2 + (y+70)^2 \leq R^2 \\ 0 & otherwise \end{cases}, \qquad (16)$$

$$P_\gamma(x, y) = \begin{cases} 1 & (x-70)^2 + (y-70)^2 \leq R^2 \\ 0 & otherwise \end{cases}, \qquad (17)$$

$$P_\delta(x, y) = \begin{cases} 1 & (x+70)^2 + (y-70)^2 \leq R^2 \\ 0 & otherwise \end{cases}. \qquad (18)$$

As pointed out before and by means of the pupil functions, the image functions of spots could be calculated as follow

$$I_\beta = |FFT(P_\beta \exp(-i(\varphi + Q(-z_2 - z_3))))|^2, \qquad (19)$$

$$I_\alpha = |FFT(P_\alpha \exp(-i(\varphi + Q(+z_2 - z_3))))|^2, \qquad (20)$$

$$I_\gamma = |FFT(P_\gamma \exp(-i(\varphi + Q(-z_2 + z_3))))|^2, \qquad (21)$$

$$I_\delta = |FFT(P_\delta \exp(-i(\varphi + Q(+z_2 + z_3))))|^2, \qquad (22)$$

where $z_2$ and $z_3$ are the Zernike tilt terms and $Q$ is the tilt coefficient. In our simulation we used Harding et al.'s MATLAB source code to simulate a phase screen with Kolmogrov statistics using interpolative methods, which produces a Kolmogrov phase screen in the desired size [14]. The resultant image of four spots could be calculated by (see figure 4)

$$I = I_\alpha + I_\beta + I_\gamma + I_\delta. \tag{23}$$

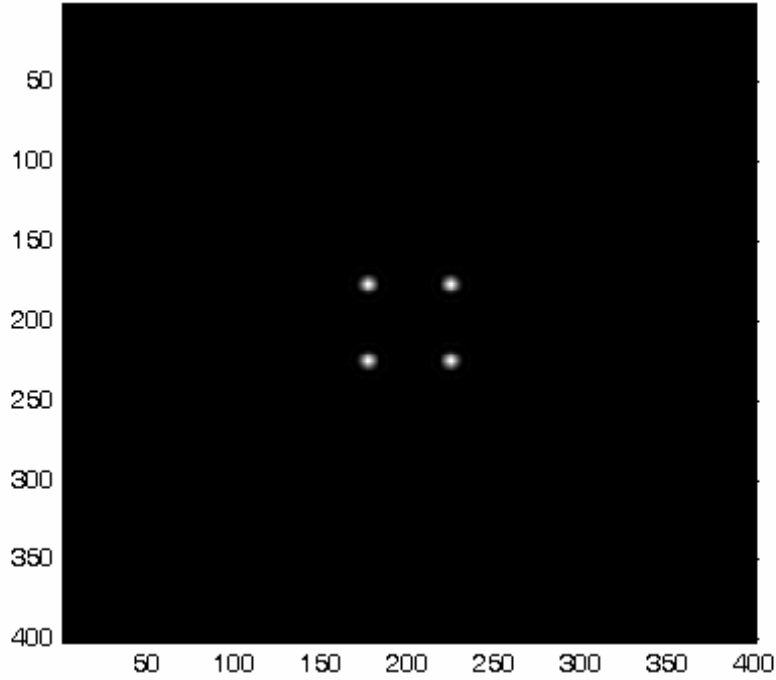

Figure 4. Image of simulated four spots.

*3.2 Primary aberration measurement by the 4- aperture DIMM*
In the theory of Hartmann test with 4-aperture screen, it is generally assumed that the higher order aberrations are negligible. However, the atmospheric high order aberrations must be taken into account in the measurement of the primary aberrations by 4-aperture DIMM. To set out our work, we generate three different sets of atmospheric phase screen by Harding et al.'s code [14]. Each set has 200 samples. To study the effect of the higher order aberrations, we decompose the generated phase screens into the Zernike modes and generate the new sets of the phase screens. Each new phase screen includes 8 modes of Zernike aberrations. In fact, in these three new sets, we cut high order aberrations from original sets. Thus, one may compare the calculated coefficients obtained from the original phase screen with those of the new phase screen with 8 Zernike modes. Using the method introduced in the subsection 3.1, we simulate 4-spot images then we calculate each spot centroid.

It should be noted that in the DIMM data analysis, measurements of the absolute local tilts of the spots are not that much desired because of the telescope tracking error or wind effect on the telescope. As pointed out in section 2, for measurement of the primary aberrations, we just need the x and y components of the spot displacements from the ideal spots positions. Also, the errors in tilt component measurements do not affect the other primary aberration measurements.

If no local tilts exist, the average coordinates of the real four spot centroids are located exactly at the position of the average coordinates of the ideal four spot centroids. Therefore, if we locate the average coordinates of the ideal four spot centroids at the average coordinates of the real four spot centroids position and then reconstruct the ideal four spot by the measured distance for each ideal spot, we could ignore tilt terms and determine the transverse aberrations for four spots.

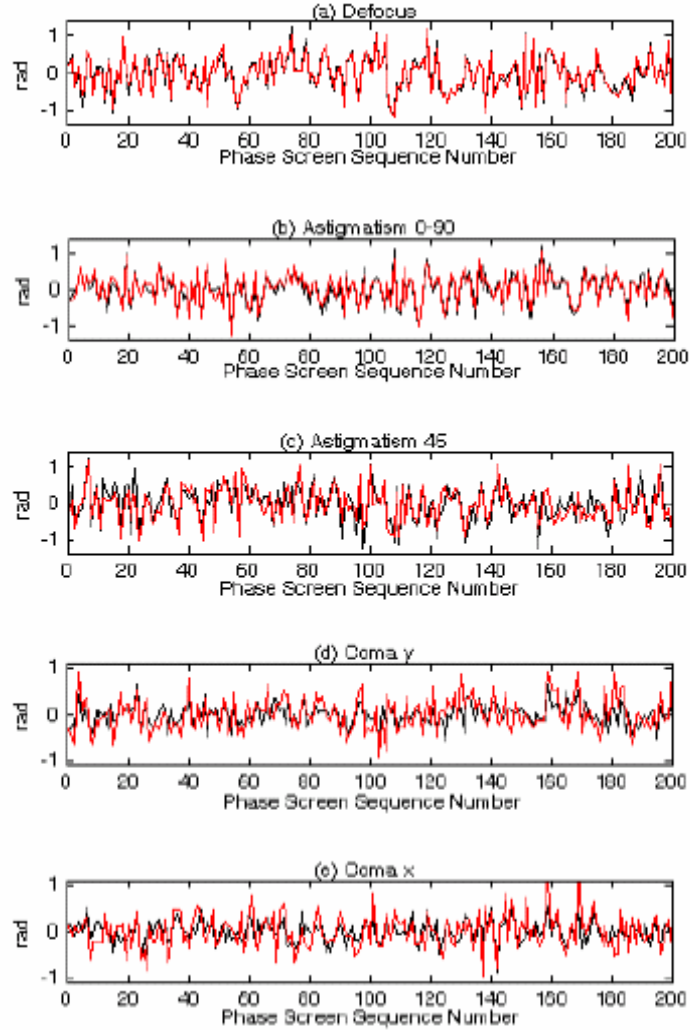

Figure 5. Primary aberrations for one set. Black curves show the calculated coefficients by 4-aperture DIMM for phase screens set which have 8 modes of Zernike aberrations in radian and red curves show the calculated coefficients by 4-aperture DIMM for original phase screens in radian. (a) Defocus (b) Astigmatism 0 or 90 (c) Astigmatism $\pm 45$ (d) Coma in y direction (e) Coma in x direction.

Using Eqs. 6-10, one may obtain five aberration coefficients the substitution of which in Eq.1 helps to reconstruct the phase screens. These phase screens are then decomposed into the Zernike modes of aberrations. The Zernike coefficients for calculated aberrations by the 4-aperture DIMM for phase screens with 8 modes of aberration together with that of the original phase distribution shows the agreement for defocus and astigmatism terms (figure 5). We calculate initial phase screen variances for the defocus, astigmatism, and coma terms. Then, we compare them with the variances of calculated Zernike terms. Results are shown in figure 6. The horizontal axis is the initial phase screen variance for a set of 200 phase screen samples and the vertical axis is the calculated variance using our method. As illustrated in this figure, our calculated variances by the 4-aperture DIMM using the phase screens with 8 modes of aberrations are in agreement with initial phase variances for all primary aberrations. However, the calculated variances for the original phase screen sets agree with the initial phase variances only for the defocus and astigmatism terms.

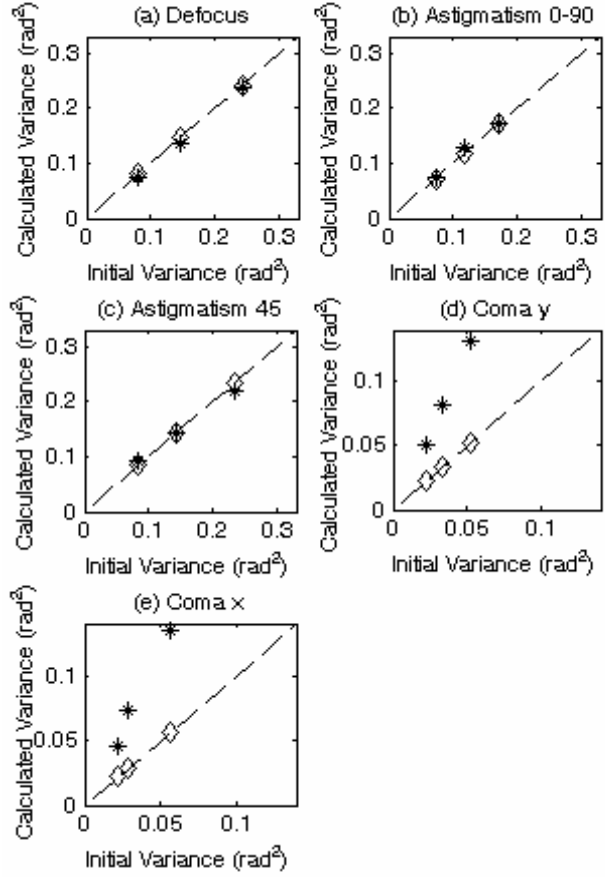

Figure 6. Horizontal axis is initial phase screen variance ($rad^2$) and the vertical axis is the calculated variance ($rad^2$) by 4-aperture DIMM. Stars show the original sets. Diamonds show the set that each phase screen has 8 modes aberrations. (a) Defocus (b) Astigmatism 0 or 90 (c) Astigmatism $\pm$ 45 (d) Coma in y direction (e) Coma in x direction.

It seems that the discrepancy in the measurement of the coma terms is due to the displacement of the centroid which is caused by the atmospheric higher order aberrations. The major contributions of higher order aberrations are Trefoil terms ($z_9, z_{10}$), which have the same expected variances as those of the coma terms and are about 4 times less than those of the defocus and astigmatism terms [15]. After these terms, the higher order aberrations in the largest case have the expected variances about 3.5 times less than that of the coma terms and 10 times less than expected variances for the defocus and astigmatism terms [15]. They could be assumed to be negligible for measuring defocus and astigmatism.

In order to quantify the percentage of the error between the predefined wave front and the reconstructed ones by the 4-aperture DIMM, the following equation is used

$$ER = \sqrt{\iint [\varphi_{4DIMM} - \varphi_i]^2 dA} \Big/ \sqrt{\iint \varphi_i^2 dA}, \qquad (24)$$

where $\varphi_{4DIMM}$ is the phase distribution which is reconstructed by the 4-aperture DIMM, $\varphi_i$ is the initial phase distribution and $dA$ is the element of telescope aperture area. In figure 7, histograms of the calculated *ER* for the collection of all three sets (600 samples) are plotted. The horizontal axis is the *ER* and the vertical axis is the number of phase screens. To have a convenient scale in figure 7, we cumulate the *ER* values which are greater than 4 at the *ER=4.5*. As illustrated in figure 7, in measurement of the defocus, astigmatism 0 or 90, astigmatism 45, coma in x direction and coma in y direction, 73%, 69%, 48%, 30% and 27% of data have *ER* less than 0.5 and 86%, 83%, 72%, 46% and 47% of data have *ER* less than 1, respectively.

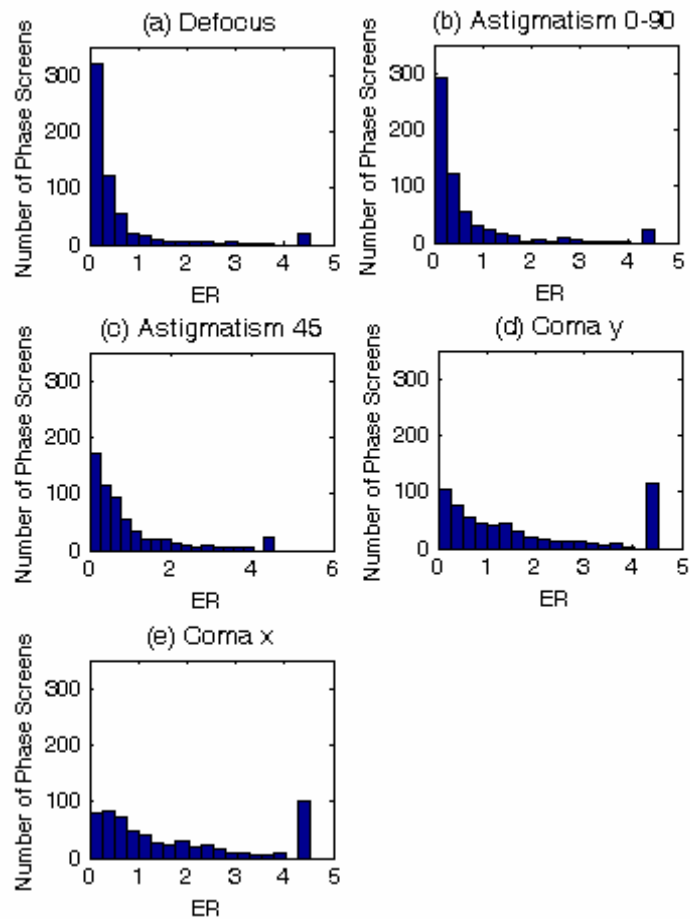

Figure 7. Horizontal axis is ER and vertical axis is the number of phase screens. (a) Defocus (b) Astigmatism 0 or 90 (c) Astigmatism $\pm 45$ (d) Coma in y direction (e) Coma in x direction.

*3.3 Determination of the ideal images of the four spots*
In contrast to the calculation of the defocusing, astigmatism and coma terms, the tilt terms cannot be calculated accurately because of the telescope tracking error. As discussed earlier, the average coordinates of the real four spot centroids is not shifted by the presence of defocusing, astigmatism, coma (by added term $-d'^2$), so by ignoring the tilt terms, we can assume the average coordinates of the real four spot centroids of each image is located exactly in the average coordinates of the ideal four spot centroids.
The ideal images of the four spots can be obtained by two methods:
  1- Pointing the telescope on a faint star and then taking an image the exposure time of which must be long enough to average out the turbulence effects but short enough to avoid any degradation that is due to telescope tracking errors.
  2- Averaging on the centroid of a large number of short exposure-time images to eliminate the seeing effects and then finding the position of the ideal four spots.

**4. Conclusion**
The main focus of the present study was to measure the atmospheric primary aberrations in the presence of the Hartmann test with four apertures. By means of the numerical simulations, we showed that this method is able to measure the defocus and the astigmatism aberrations. In the course of our exploration, a purposeful modification was made in the DIMM method data analysis so that in addition to the Fried parameter, determination of the three more primary aberrations of atmosphere, i.e., defocusing, astigmatisms became possible. However, the results obtained by this method were not reliable for the coma aberration. The evidential supports of the study show that through applying the 4-aperture DIMM, one may simply calculate the coherence time obtained by defocusing measurements.

**Acknowledgement**
The authors gratefully acknowledge *C M Harding, R A Johnston and R G Lane* for working out the MATLAB source code to simulate the phase screen with Kolmogrov statistics and Marcos Van Dam for preparing us this source code.